\documentclass[english,aps,tightenlines,showpacs, showkeys, notitlepage]{revtex4-2}
\usepackage[T1]{fontenc}
\usepackage[latin9]{inputenc}
\usepackage{color}
\usepackage{babel}
\usepackage{amsmath}
\usepackage{amssymb}
\usepackage[unicode=true,pdfusetitle,
 bookmarks=true,bookmarksnumbered=false,bookmarksopen=false,
 breaklinks=false,pdfborder={0 0 0},pdfborderstyle={},backref=false,colorlinks=true]
 {hyperref}
\begin{document}
\title{Cosmological application of the Maxwell gravity}
\author{Salih Kibaro\u{g}lu}
\email{salihkibaroglu@maltepe.edu.tr}

\date{\today}
\begin{abstract}
In this study, we consider a cosmological model for the Maxwell gravity
which is constructed by gauging the semi-simple extended Poincaré
algebra. Inspired by the Einstein-Yang-Mills theory, we describe the
Maxwell gauge field in terms of two additional time-dependent scalar
fields. Within the context of a homogeneous and isotropic Friedmann-Lemaître-Robertson-Walker
universe, we derive the Friedmann equations together with new contributions.
Additionally, we find the exact expressions for the additional scalar
fields, considering the exponential evolution of the scale factor.
Moreover, we investigate the gauge theory of gravity based on the
Maxwell algebra and demonstrate that this model leads to the (anti)
de Sitter universe as well as a non-accelerated universe model.
\end{abstract}
\affiliation{Maltepe University, Faculty of Engineering and Natural Sciences, 34857,
Istanbul, Turkey}
\affiliation{Institute of Space Sciences (IEEC-CSIC) C. Can Magrans s/n, 08193
Barcelona, Spain}
\maketitle

\section{Introduction}

Einstein's general theory of relativity is widely regarded as the
most effective theoretical framework for describing gravitational
phenomena in a large range of scales. It has been tested through many
experiments and observations. However, despite its remarkable observational
accomplishments, several unresolved puzzles remain. One such example
is the nature of dark energy, which is thought to be responsible for
the accelerated expansion of the universe still remains unknown. Additionally,
the theory is not useful at very high energies where quantum effects
are expected to become important. These considerations constitute
the driving force for examining generalized theories of gravity.

There exists an interesting extended gravitation theory that comes
from the gauge theory of the Maxwell algebra. The Maxwell algebra
can be interpreted as a modification of the Poincaré algebra by six
additional tensorial Abelian generators that make the four-momenta
non-commutative $\left[P_{a},P_{b}\right]=iZ_{ab}$ \citep{bacry1970group,schrader1972maxwell,soroka2005tensor}.
If one constructs a gauge theory of gravity based on this algebra,
it leads to a generalized theory of gravity that includes the cosmological
constant and an additional term to the energy-momentum tensor and
this gravitational model is called the Maxwell gravity or the Maxwell
gauge theory of gravity \citep{azcarraga2011generalized,durka2011gauged,soroka2012gauge,azcarraga2014minimalsuper,cebeciouglu2014gauge,cebeciouglu2015maxwell,concha2015generalized,kibarouglu2019maxwellSpecial,kibarouglu2019super,kibarouglu2020generalizedConformal,kibarouglu2021gaugeAdS,cebeciouglu2021maxwellMetricAffine}.
Up to now, this energy-momentum term has not been extensively analyzed
yet, but it is known that such an additional term may be related to
dark energy \citep{frieman2008dark,padmanabhan2009dark}. In the cosmological
framework, a minimal cosmological model related to this symmetry can
be found in \citep{durka2011local} with a small discussion. Besides,
it is also discussed that the gauge fields of the Maxwell symmetry
may provide a geometric background to describe vector inflatons in
cosmological models \citep{Azcarraga2013maxwellApplication}. For
the non-gravitational case, this symmetry group is used to describe
a particle moving in a Minkowski spacetime filled with a constant
electromagnetic background field which is formed by the additional
degrees of freedom related to $Z_{ab}$. From this idea, Maxwell symmetry
is considered as the symmetry group of a particle moving in a constant
electromagnetic field \citep{Gomis:2009Deformations,bonanos2009infinite}.
It is also used to describe planar dynamics of the Landau problem
\citep{fedoruk2012new}, higher spin fields \citep{fedoruk2013maxwell,fedoruk2013new},
and applied to the string theory as an internal symmetry of the matter
gauge fields \citep{hoseinzadeh20142+}.

The gauge field configurations in the context of the Friedmann-Lemaître-Robertson-Walker
(FLRW) cosmology have been discussed in the literature before \citep{cervero1978classical,henneaux1982remarks,galt1991yang,moniz1993dynamics,bamba2008inflationary,gal2008non,maleknejad2013gauge,guarnizo2020dynamical}.
For example, in the Yang-Mills (YM) theories in which a non-Abelian
YM field couples to the scalar curvature, one examines the cosmological
consequences of the nonminimal gravitational coupling of the YM field.
YM theory is also widely used in inflationary cosmology \citep{maleknejad2011non,sheikh2012gauge,maleknejad2013gauge,bielefeld2015cosmological}
(for a complete review see \citep{maleknejad2013gauge_report}) and
used in the models for dark energy \citep{Zhao:2005TheStateEquation,gal2008non,bamba2008inflationary,gal2012yang,elizalde2013cosmological,setare2013warm,rinaldi2015dark,mehrabi2017gaugessence}
and dark matter \citep{zhang2010dark,buen2015non,gross2015non}. Our
aim in this paper is to find the potential effects of the additional
energy-momentum tensor produced by the Maxwell gravity on the evolution
of the universe. For this purpose, inspired by the YM theory and gauge
field construction method proposed by \citep{Azcarraga2013maxwellApplication},
we study to construct a cosmological model for a class of the Maxwell
gravity which is obtained gauging the semi-simple extended Poincaré
algebra \citep{Soroka:2009SemiSimpleSuper,Gomis:2009Deformations,soroka2012gauge}.

This paper is organized as follows. In Section II, we provide a concise
overview of gauging the semi-simple extended Poincaré algebra. In
Section III, we investigate two scenarios within a homogeneous and
isotropic FLRW universe to derive the Friedman equations for the Maxwell
gravity. Finally, the last section concludes the paper by giving some
discussion.

\section{Gauge theory of the semi-simple extended Poincaré group\label{sec:Gauge-theory-of}}

The Maxwell algebra \citep{soroka2005tensor} is a non-central extension
of the Poincaré algebra which has non-commutative momentum generators
as $\left[P_{a},P_{b}\right]=i\lambda Z_{ab}$. Here $Z_{ab}$ is
the additional antisymmetric tensorial generator which has an Abelian
characteristic. One can derive the Maxwell algebra by applying the
Lie algebra expansion method outlined in \citep{Izaurieta:2006ExpandingLie,Salgado:2014soD1}.
from the (A)dS algebra. This technique makes it possible to generate
two sets of algebras known as generalized Poincaré algebras (also
referred to as $\mathfrak{B}_{n}$ algebra, where the Maxwell algebra
is a specific instance called $\mathfrak{B}_{4}$ algebra) and generalized
AdS algebras (which is known as $\text{AdS}\mathcal{L}_{n}$ algebras).

In this section, we shortly review the gauge theory of gravity based
on the semi-simple extended Poincaré algebra \citep{Soroka:2009SemiSimpleSuper,soroka2012gauge}
which corresponds to $\text{AdS}\mathcal{L}_{4}$ algebra \citep{Cardenas:2022GeneralizedEinstein}
and it can be established by direct sum of the anti-de Sitter AdS
algebra $so\left(3,2\right)$ and the Lorentz algebra $so\left(3,1\right)$
\citep{Soroka:2009SemiSimpleSuper}. This algebra extends the Poincare
algebra by using a non-Abelian tensorial generator $Z_{ab}$ and can
be seen as the modification of the Maxwell algebra \citep{Gomis:2009Deformations}.
The gauge theory of gravity based on this algebra lead to generalizing
the Maxwell gravity which was discussed in \citep{soroka2012gauge}
in the context of the cosmological term problem.

The Lie algebra of this extended symmetry can be given as follows,

\begin{eqnarray}
\left[M_{ab},M_{cd}\right] & = & i\left(\eta_{ad}M_{bc}+\eta_{bc}M_{ad}-\eta_{ac}M_{bd}-\eta_{bd}M_{ac}\right),\nonumber \\
\left[M_{ab},Z_{cd}\right] & = & i\left(\eta_{ad}Z_{bc}+\eta_{bc}Z_{ad}-\eta_{ac}Z_{bd}-\eta_{bd}Z_{ac}\right),\nonumber \\
\left[Z_{ab},Z_{cd}\right] & = & i\alpha\left(\eta_{ad}Z_{bc}+\eta_{bc}Z_{ad}-\eta_{ac}Z_{bd}-\eta_{bd}Z_{ac}\right),\nonumber \\
\left[P_{a},P_{b}\right] & = & i\lambda Z_{ab},\nonumber \\
\left[M_{ab},P_{c}\right] & = & i\left(\eta_{bc}P_{a}-\eta_{ac}P_{b}\right),\nonumber \\
\left[Z_{ab},P_{c}\right] & = & i\alpha\left(\eta_{bc}P_{a}-\eta_{ac}P_{b}\right),\label{eq: algebra_ss}
\end{eqnarray}
where $M_{ab}$ is the generator of rotation, $P_{a}$ is the generator
of translation, $Z_{ab}$ is the tensor generator, $\eta_{ab}$ is
the Minkowski metric which have $diag\left(\eta_{ab}\right)=\left(+,-,-,-\right)$
and the group indices $a,b,...=0,...,3$. Here, the constant $\lambda$
has the unit of $L^{-2}$ which will be related to the cosmological
constant where $L$ is considered as the unit of length and $\alpha$
is a dimensionless constant. In addition to the Poincaré algebra,
this algebra contains six new additional non-Abelian tensorial generators
$Z_{ab}$ which behave as an antisymmetric second-rank Lorentz tensor.

To construct gauge theory of gravity based on the semi-simple Poincaré
algebra, we firstly introduce the following one-form gauge field,

\begin{equation}
\mathcal{A}\left(x\right)=e^{a}P_{a}+\frac{1}{2}B^{ab}Z_{ab}-\frac{1}{2}\omega^{ab}M_{ab},\label{eq: gauge field 1}
\end{equation}
where $e^{a}\left(x\right)=e_{\mu}^{a}dx^{\mu}$, $B^{ab}\left(x\right)=B_{\mu}^{ab}dx^{\mu}$,
and $\omega^{ab}\left(x\right)=\omega_{\mu}^{ab}dx^{\mu}$ are the
one form gauge fields of corresponding generators. Also, the unit
dimension of all gauge fields have zero other than $\left[e^{a}\right]=L$.
Using the structure equation $\mathcal{F}=d\mathcal{A}+\frac{i}{2}\left[\mathcal{A},\mathcal{A}\right]$
and defining the curvatures as $\mathcal{F}\left(x\right)=F^{a}P_{a}+\frac{1}{2}F^{ab}Z_{ab}-\frac{1}{2}R^{ab}M_{ab}$,
we find the associated two-form curvatures as,

\begin{eqnarray}
F^{a} & = & \mathcal{D}e^{b}-\alpha B_{\,\,c}^{a}\wedge e^{c},\nonumber \\
F^{ab} & = & \mathcal{D}B^{ab}-\alpha B_{\,\,c}^{a}\wedge B^{cb}-\lambda e^{a}\wedge e^{b},\nonumber \\
R^{ab} & = & \mathcal{D}\omega^{ab},\label{eq: curvatures}
\end{eqnarray}
where $\mathcal{D}\Phi=[d+\omega]$$\Phi$ is the Lorentz covariant
derivative. Then, taking the covariant derivative of the curvatures,
we can obtain the Bianchi identities as follows;

\begin{eqnarray}
\mathcal{D}F^{a} & = & R_{\,\,c}^{a}\wedge e^{c}-\alpha\mathcal{D}B_{\,\,c}^{a}\wedge e^{c}+\alpha B_{\,\,c}^{a}\wedge\mathcal{D}e^{c},\nonumber \\
\mathcal{D}F^{ab} & = & R_{\,\,\,\,c}^{[a}\wedge B^{c|b]}-\alpha\mathcal{D}B_{\,\,c}^{[a}\wedge B^{cb]}-\lambda\mathcal{D}e^{[a}\wedge e^{b]},\nonumber \\
\mathcal{D}R^{ab} & = & 0.
\end{eqnarray}

At this point, it is worth noting that a shifted connection can be
established through the utilization of the one-forms $B^{ab}\left(x\right)$
to extend the Riemannian connection $\omega^{ab}\left(x\right)$ in
the subsequent manner \citep{azcarraga2011generalized,Cebecioglu:2023Maxwellf(R)},

\begin{equation}
\tilde{\omega}^{ab}=\omega^{ab}-\alpha B^{ab}.
\end{equation}
This modified connection can be construed as a generalization of the
Riemannian connection $\omega^{ab}$ to a non-Riemannian connection.
In this particular context, the antisymmetry of $B^{ab}$ indicates
that the system under consideration can be described by an Einstein--Cartan
geometry. Furthermore, the non-metricity tensor associated with this
geometry is identically zero due to the vanishing of the symmetric
component of the shifted connection, i.e., $\tilde{\omega}^{\left(ab\right)}=0$.
In this sense, it is possible to regard this model as a particular
extension to a non-Riemannian framework that is determined by the
structure of the semi-simple Poincaré algebra.

\subsection{Gravitational action}

From this background, one can construct the following action by introducing
a shifted curvature defined as $\mathcal{J}^{ab}=R^{ab}-\mu F^{ab}$,

\begin{eqnarray}
S_{g} & = & \frac{1}{4\kappa\mu\lambda}\int\mathcal{J}\wedge*\mathcal{J}\nonumber \\
 & = & \frac{1}{4\kappa}\int\left(\epsilon_{abcd}R^{ab}\wedge e^{c}\wedge e^{d}+\frac{\mu\lambda}{2}\epsilon_{abcd}e^{a}\wedge e^{b}\wedge e^{c}\wedge e^{d}\right)\nonumber \\
 &  & +\epsilon_{abcd}\left(\frac{\alpha}{\lambda}R^{ab}\wedge B_{\,\,e}^{c}\wedge B^{ed}-\mu\mathcal{D}B^{ab}\wedge e^{c}\wedge e^{d}+\frac{\alpha}{2}B_{\,\,e}^{a}\wedge B^{eb}\wedge e^{c}\wedge e^{d}\right)\nonumber \\
 &  & +\epsilon_{abcd}\left(\frac{\mu}{2\lambda}\mathcal{D}B^{ab}\wedge\mathcal{D}B^{cd}+\frac{\mu\alpha}{\lambda}\mathcal{D}B^{ab}\wedge B_{\,\,e}^{c}\wedge B^{ed}+\frac{\mu\alpha^{2}}{2\lambda}B_{\,\,e}^{a}\wedge B^{eb}\wedge B_{\,\,f}^{c}\wedge B^{fd}\right),\label{eq: action}
\end{eqnarray}
where we have neglected the total derivative terms. The first part
of the above action contains the Einstein-Hilbert term together with
the cosmological term and the other parts consist of mixed terms.
Note that the asterisk represents Hodge duality operation, $\kappa=8\pi G$
and $\mu$ is a constant. 

In order to obtain the corresponding field equations, the action can
be varied with respect to the gauge fields $\omega^{ab}\left(x\right)$,
$B^{ab}\left(x\right),$and $e^{a}\left(x\right)$, respectively.
Thus one gets the following equations of motion,

\begin{eqnarray}
\epsilon_{abcd}\mathcal{D}\mathcal{J}^{ab}+\mu\epsilon_{abcd}\mathcal{J}_{\,\,e}^{[a}\wedge B^{e|b]} & = & 0,\label{eq: var_w_B}\\
\epsilon_{abcd}\mathcal{J}^{ab}\wedge e^{c} & = & 0,\label{eq: EFE1}
\end{eqnarray}
Here, we want to note that the variation $\omega^{ab}\left(x\right)$
and $B^{ab}\left(x\right)$ satisfy the same equation (\ref{eq: var_w_B}). 

Using Eq.(\ref{eq: EFE1}) and transforming the tangent indices to
world space-time indices, one can show that the generalized Einstein
field equations can be written as follows,

\begin{equation}
R_{\,\,\alpha}^{\mu}-\frac{1}{2}\delta_{\alpha}^{\mu}R=T{}_{\,\,\alpha}^{\mu},\label{eq: einstein_filed_eq-2}
\end{equation}
where the Greek indices $\mu,\nu,...=0,...,3$ and $T_{\,\,\alpha}^{\mu}$
represents the Maxwell symmetry contributions which can be shown as,

\begin{equation}
T_{\,\,\alpha}^{\mu}=\mu\left[e_{a}^{\mu}e_{b}^{\beta}\left(\mathcal{D}_{[\alpha}B_{\beta]}^{ab}-\alpha B_{[\alpha\,\,c}^{a}B_{\beta]}^{cb}\right)-\frac{1}{2}\delta_{\alpha}^{\mu}e_{a}^{\rho}e_{b}^{\sigma}\left(\mathcal{D}_{[\rho}B_{\sigma]}^{ab}-\alpha B_{[\rho\,\,c}^{a}B_{\sigma]}^{cb}\right)+3\delta_{\alpha}^{\mu}\lambda\right].\label{eq: EM_tot}
\end{equation}
We also assume that $T_{\,\,\alpha}^{\mu}$ has a perfect fluid characteristics
as $T_{\,\,\alpha}^{\mu}=diag\left(\rho,-P,-P,-P\right)$ with the
energy density $\rho$$\left(t\right)$ and pressure $P\left(t\right)$.
So the all matter-energy content of the universe is described by the
Maxwell curvature contributions in this formulation. 

On the other hand, the term $3\delta_{\alpha}^{\mu}\lambda$ can be
considered as a cosmological constant when we multiply with the parameter
$\mu$. So we can divide the Eq.(\ref{eq: EM_tot}) into two parts
such as 
\begin{equation}
T_{\,\,\alpha}^{\mu}=T_{B}{}_{\,\,\alpha}^{\mu}+T_{\Lambda}{}_{\,\,\alpha}^{\mu},
\end{equation}
where $T_{B}{}_{\,\,\alpha}^{\mu}$ and $T_{\Lambda}{}_{\,\,\alpha}^{\mu}$
represents the energy-momentum tensors of the Maxwell gauge field
$B_{\mu}^{ab}$ and the cosmological constant $\Lambda=\lambda\mu$.
Thus we can write the field equation in Eq.(\ref{eq: einstein_filed_eq-2})
as the following form,

\begin{equation}
R_{\,\,\alpha}^{\mu}-\frac{1}{2}\delta_{\alpha}^{\mu}R-3\Lambda\delta_{\alpha}^{\mu}=T_{B}{}_{\,\,\alpha}^{\mu},\label{eq: einstein_filed_eq}
\end{equation}
where we have used the explicit expression $T_{\Lambda}{}_{\,\,\alpha}^{\mu}=3\Lambda\delta_{\alpha}^{\mu}$
to indicate the cosmological constant in a more clear formulation.
We assume again that $T_{B}{}_{\,\,\alpha}^{\mu}$ and $T_{\Lambda}{}_{\,\,\alpha}^{\mu}$
represent perfect fluids and thus the total energy density and total
pressure take the following forms,

\begin{equation}
\rho=\rho_{B}+\rho_{\Lambda},\label{eq: rho_total}
\end{equation}
and
\begin{equation}
P=P_{B}+P_{\Lambda},\label{eq: P_total}
\end{equation}
where we assume that all energy densities are positive definite. In
the light of this result, we can say that the standard gravitational
equation is extended to include a cosmological constant and additional
energy-momentum tensors in terms of the Maxwell symmetry. These are
the main characteristics of the Maxwell extended (super)-gravitational
theories. In the next section, we will analyze the cosmological behavior
of this gravitational equation.

\section{Cosmological setup\label{sec:Cosmological-setup}}

Motivated by the cosmological implications of the Einstein-Yang-Mills
framework, where a non-Abelian gauge field was utilized to explore
inflationary or late-time accelerated expansion, we consider finding
potential cosmological consequences of the Maxwell gauge fields $B_{\mu}^{ab}$
which present in gravitational field equation in Eq.(\ref{eq: einstein_filed_eq}).
In general, for most cosmological models, the metric is assumed to
be a flat FLRW metric because the universe seems to be nearly flat
so we begin our analysis with the following line element,

\begin{eqnarray}
ds^{2} & = & dt^{2}-a\left(t\right)^{2}\left(dx^{2}+dy^{2}+dz^{2}\right).\label{eq: FLRW metric}
\end{eqnarray}
In this space, the space-time metric can be defined in terms of the
vierbein fields $e_{\mu}^{a}\left(x\right)$,

\begin{equation}
g_{\mu\nu}\left(x\right)=\eta_{ab}e_{\mu}^{a}e_{\nu}^{b},
\end{equation}
where $\eta_{ab}$ is the tangent space metric which is
\begin{equation}
\eta_{ab}=g_{\mu\nu}\left(x\right)e_{a}^{\mu}e_{b}^{\nu}.
\end{equation}
Here the vierbein field can be defined as,
\begin{equation}
e_{\mu}^{a}\left(x\right)=\left(e_{0}^{a},e_{i}^{a}\right)=\left(\delta_{0}^{a},-\delta_{i}^{a}a\left(t\right)\right),\label{eq: vierbein}
\end{equation}
and satisfy the following relations
\begin{equation}
e_{\mu}^{a}\left(x\right)e_{b}^{\mu}\left(x\right)=\delta_{b}^{a},\,\,\,\,\,\,\,\,\,\,\,\,e_{a}^{\mu}\left(x\right)e_{\nu}^{a}\left(x\right)=\delta_{\nu}^{\mu},
\end{equation}
with

\begin{equation}
e=det\left(e_{\mu}^{a}\right)=\sqrt{-det\left(g_{\mu\nu}\right)}.
\end{equation}

In the conventional non-Abelian gauge theory which is minimally coupled
to Einstein's gravity in four dimensions, one can consider a simple
gauge invariant Lagrangian such as 
\begin{equation}
\mathcal{L}=\frac{1}{2}R+\mathcal{L}_{G}\left(F_{\mu\nu}^{A}\right)
\end{equation}
where $\mathcal{L}_{G}\left(F_{\mu\nu}^{A}\right)$ is a generic gauge
invariant action which may include the Yang-Mills term made out of
powers of the field strength $F_{\mu\nu}^{A}$ and $A,B,...=1,2,...\text{dim}\mathcal{G}$.
In general, we note that the gauge group $\mathcal{G}$ can be any
non-Abelian compact group. Here the field strength tensor is defined
as
\begin{equation}
F_{\mu\nu}^{A}=\partial_{\mu}A_{\nu}^{A}-\partial_{\nu}A_{\mu}^{A}-gf_{\,\,BC}^{A}A_{\mu}^{B}A_{\nu}^{C},
\end{equation}
where $A_{\mu}^{A}$ is the gauge field, $g$ is the gauge coupling
and $f_{\,\,BC}^{A}$ are the structure constants of the gauge group.
If we choose the gauge group $\mathcal{G}$ to be $SU\left(2\right)$,
the gauge field can be introduced as follows,
\begin{equation}
A_{\mu}^{a}=\left\{ 0,\psi\left(t\right)e_{i}^{a}\right\} ,\label{eq: gauge_field_su(2)}
\end{equation}
where $\psi\left(t\right)$ is a scalar field under rotations and
$e_{i}^{a}=-a\left(t\right)\delta_{i}^{a}$ are the triads of the
spatial hyper-surface (for more details, see the report \citep{maleknejad2013gauge_report}).

In our scenario, the formulation of the Maxwell gauge field slightly
differs from the expression given in Eq.(\ref{eq: gauge_field_su(2)}).
According to \citep{Azcarraga2013maxwellApplication}, we define the
Maxwell gauge fields $B_{\mu}^{ab}$ using one-dimensional fields
$\psi\left(t\right)$ and $\zeta\left(t\right)$,

\begin{equation}
B_{\mu}^{0s}\left(x\right)\rightarrow B_{\mu}^{0s}\left(t\right)=\left(0,\delta_{i}^{s}\psi\left(t\right)\right),\label{eq: B_1}
\end{equation}

\begin{equation}
B_{\mu}^{rs}\left(x\right)\rightarrow B_{\mu}^{rs}\left(t\right)=\left(0,\epsilon_{i}^{rs}\zeta\left(t\right)\right),\label{eq: B_2}
\end{equation}
where $r,s=1,2,3$ are the tangent indices and $i,j=1,2,3$ are the
space-time indices. Here, $\delta_{i}^{s}$ and $\epsilon_{i}^{rs}$
are three dimensional $so\left(3\right)$ tensors.

\subsection{$\alpha=\mu$ case}

In this case, we interested in the torsion-free condition which requires
the condition $e^{a}\wedge e_{e}=\frac{\mu}{\lambda}B_{\,\,f}^{a}\wedge B_{\,\,e}^{f}$
to satisfy the equations of motion in Eq.(\ref{eq: var_w_B}). Therefore,
the spin connection $\omega_{\mu}^{ab}\left(x\right)$ can be expressed
in terms of vierbein and the Levi Civita connection $\Gamma_{\mu\sigma}^{\nu}$
as

\begin{equation}
\omega_{\mu}^{ab}=e_{\nu}^{a}\partial_{\mu}e^{b\,\nu}+e_{\nu}^{a}\Gamma_{\mu\sigma}^{\nu}e^{b\,\sigma}.\label{eq: spin_connection}
\end{equation}
Then applying Eqs.(\ref{eq: B_1}), (\ref{eq: B_2}) and (\ref{eq: spin_connection})
to the field equation in Eq.(\ref{eq: einstein_filed_eq}), we are
ready to find the Friedmann equations. The first equation can be derived
from the $\left(0,0\right)$ component of Eq.(\ref{eq: einstein_filed_eq}),

\begin{equation}
\left(\frac{\dot{a}}{a}\right)^{2}=\frac{\mu}{a^{2}}\left[2\dot{a}\psi-\mu\left(\psi^{2}-\zeta^{2}\right)\right]+\Lambda,\label{eq: Friedmann_00}
\end{equation}
where the dot denotes derivative with respect to the cosmic time $t$.
The $\left(i,i\right)$ component of Eq.(\ref{eq: einstein_filed_eq})
leads to the following equation
\begin{equation}
\frac{2\ddot{a}}{a}+\left(\frac{\dot{a}}{a}\right)^{2}=\frac{\mu}{a^{2}}\left[2\dot{a}\psi+2a\dot{\psi}-\mu\left(\psi^{2}-\zeta^{2}\right)\right]+3\Lambda,\label{eq: Friedmann_ii}
\end{equation}
Then by making use of Eqs.(\ref{eq: Friedmann_00}) and (\ref{eq: Friedmann_ii}),
the acceleration equation can be obtained as,

\begin{equation}
\frac{\ddot{a}}{a}=\frac{\mu\dot{\psi}}{a}+\Lambda,\label{eq: Friedmann_acceleration}
\end{equation}
Here according to this equation, we see that the acceleration of the
universe depends on $\psi\left(t\right)$ and $\Lambda$. If $\psi\left(t\right)$
is constant or zero the corresponding universe exhibits behavior like
the ordinary de Sitter universe in which the acceleration arises due
to the cosmological constant. On the other hand, when $\dot{\psi}=-a\Lambda/\mu$,
the acceleration vanishes. 

Moreover, by using the Eqs.(\ref{eq: Friedmann_00}) and (\ref{eq: Friedmann_ii})
one can find the energy density and the pressure expressions for the
Maxwell gauge field and the cosmological constant as follows,

\begin{eqnarray}
\rho_{B} & = & \frac{3\mu}{\kappa a^{2}}\left[2\dot{a}\psi-\mu\left(\psi^{2}-\zeta^{2}\right)\right],\label{eq: rho_P_B-m}
\end{eqnarray}

\begin{eqnarray}
P_{B} & = & -\frac{\mu}{\kappa a^{2}}\left(2\dot{a}\psi+2a\dot{\psi}-\mu\left(\psi^{2}-\zeta^{2}\right)\right)\nonumber \\
 & = & -\frac{2\mu\dot{\psi}}{\kappa a}-\frac{1}{3}\rho_{B},
\end{eqnarray}

\begin{equation}
\rho_{\Lambda}=\frac{3\Lambda}{\kappa},\,\,\,\,\,\,P_{\Lambda}=-\frac{3\Lambda}{\kappa}.\label{eq: rho_P_Lambda-m}
\end{equation}

For completeness, if we combine Eqs.(\ref{eq: Friedmann_00}) and
(\ref{eq: Friedmann_acceleration}), we also derive the time derivative
of the Hubble parameter ($H\left(t\right)=\frac{\dot{a}\left(t\right)}{a\left(t\right)}$)
as
\begin{equation}
\dot{H}=\frac{\mu}{a^{2}}\left(a\dot{\psi}-2\dot{a}\psi+\mu\left(\psi^{2}-\zeta^{2}\right)\right)=-\frac{\kappa}{2}\left(\rho_{B}+P_{B}\right),\label{eq: der_H}
\end{equation}
and taking the time derivation of Eq.(\ref{eq: Friedmann_00}) and
using Eq.(\ref{eq: der_H}), we get 
\begin{equation}
\dot{\rho}_{B}+3H\left(\rho_{B}+P_{B}\right)=0,
\end{equation}
which is the energy conservation equation for the Maxwell gauge field
contribution.

\subsubsection*{A solution for the Friedmann equations}

In the context of the exponential evolution of the scale factor, we
proceed to analyze the Friedmann equations of this model. Assuming
the scale factor takes the form $a\left(t\right)=a_{0}e^{ht}$ ($a_{0}$
and $h$ are constants) for the spatially-flat FLRW universe (\ref{eq: FLRW metric})
which is a solution of Eq.(\ref{eq: einstein_filed_eq}). Building
upon this background, we begin by solving the acceleration equation
in Eq.(\ref{eq: Friedmann_acceleration}) with respect to $\psi\left(t\right)$,
we get

\begin{equation}
\psi\left(t\right)=\frac{a_{0}e^{ht}\left(h^{2}-\Lambda\right)}{h\mu}+C,\label{eq: psi_1}
\end{equation}
and putting this equation into Eq.(\ref{eq: Friedmann_00}) or Eq.(\ref{eq: Friedmann_ii}),
one can get the following solution for $\zeta\left(t\right)$,
\begin{equation}
\zeta\left(t\right)=\pm\frac{1}{\mu h}\sqrt{a_{0}^{2}\Lambda e^{2ht}\left(\Lambda-h^{2}\right)-2\Lambda h\mu a_{0}Ce^{ht}+h^{2}\mu^{2}C^{2}}.\label{eq: zeta_1}
\end{equation}

On the other hand, if we solve Eq.(\ref{eq: Friedmann_00}) with respect
to $\psi$, we get;
\begin{equation}
\psi\left(t\right)=\frac{1}{\mu}\left(a_{0}he^{ht}\pm\sqrt{\Lambda a_{0}^{2}e^{2ht}+\mu^{2}\zeta\left(t\right)^{2}}\right),\label{eq: psi_2}
\end{equation}
then equate with Eq.(\ref{eq: psi_1}) and solve with respect to $\zeta\left(t\right)$,
we reach the same solution as given in Eq.(\ref{eq: zeta_1}). Thus
these results show the consistency of the Friedmann equations in Eqs.(\ref{eq: Friedmann_00}),
(\ref{eq: Friedmann_ii}) and (\ref{eq: Friedmann_acceleration})
under the exponential evolution of the scale factor. So Eqs.(\ref{eq: psi_1})
and (\ref{eq: zeta_1}) together with the definition of the scale
factor turn out to be a solution to the Friedmann equations. 

Furthermore, by setting the parameter $h=\sqrt{\varLambda}$ and fixing
the integration constant $C$ to zero in Eq.(\ref{eq: psi_1}), the
acceleration equations (\ref{eq: Friedmann_acceleration}) can be
transformed into those describing the pure de Sitter universe. We
also note that utilizing a similar approach, exact formulations for
$\psi\left(t\right)$ and $\zeta\left(t\right)$ can be derived in
a more comprehensive scenario where $a\left(t\right)=a_{0}e^{f(t)}$
with $f\left(t\right)$ representing an arbitrary function of the
cosmic time.

\subsection{$\alpha=0$ case}

If we apply this condition to Eq.(\ref{eq: algebra_ss}), this means
that we are working in the Maxwell algebra background with the following
commutation relations \citep{soroka2005tensor,azcarraga2011generalized},

\begin{eqnarray}
\left[M_{ab},M_{cd}\right] & = & i\left(\eta_{ad}M_{bc}+\eta_{bc}M_{ad}-\eta_{ac}M_{bd}-\eta_{bd}M_{ac}\right),\nonumber \\
\left[M_{ab},Z_{cd}\right] & = & i\left(\eta_{ad}Z_{bc}+\eta_{bc}Z_{ad}-\eta_{ac}Z_{bd}-\eta_{bd}Z_{ac}\right),\nonumber \\
\left[P_{a},P_{b}\right] & = & i\lambda Z_{ab},\nonumber \\
\left[M_{ab},P_{c}\right] & = & i\left(\eta_{bc}P_{a}-\eta_{ac}P_{b}\right).
\end{eqnarray}
Now the additional tensor generator $Z_{ab}$ exhibits Abelian characteristics.
Gauging this algebra lead to a generalized theory of gravity known
as Maxwell gravity which was suggested as a potential alternative
approach to addressing the cosmological constant problem \citep{azcarraga2011generalized}. 

After following the same gauging procedure given in Section (\ref{sec:Gauge-theory-of}),
one can derive that the curvature two-forms reduces to,
\begin{eqnarray}
F^{a} & = & \mathcal{D}e^{b},\nonumber \\
F^{ab} & = & \mathcal{D}B^{ab}-\lambda e^{a}\wedge e^{b},\nonumber \\
R^{ab} & = & \mathcal{D}\omega^{ab},\label{eq: curvatures-simple}
\end{eqnarray}
Then taking account of the action in Eq.(\ref{eq: action}) by using
Eq.(\ref{eq: curvatures-simple}), the equations of motion with respect
to the gauge fields $\omega^{ab}\left(x\right)$, $B^{ab}\left(x\right),$and
$e^{a}\left(x\right)$, takes the following form

\begin{eqnarray}
\epsilon_{abcd}\mathcal{D}\mathcal{J}^{ab}+\mu\epsilon_{abcd}\mathcal{J}_{\,\,e}^{[a}\wedge B^{e|b]} & = & 0,\nonumber \\
\epsilon_{abcd}\mathcal{D}\mathcal{J}^{ab} & = & 0,\nonumber \\
\epsilon_{abcd}\mathcal{J}^{ab}\wedge e^{c} & = & 0.\label{eq: eom_simple}
\end{eqnarray}
Finally, using the $e^{a}\left(x\right)$, one can find the field
equation similar to the formulation of Eq.(\ref{eq: EFE1}) with the
following energy-momentum tensor,
\begin{equation}
T_{B}{}_{\,\,\alpha}^{\mu}=\mu\left(e_{a}^{\mu}e_{b}^{\beta}\mathcal{D}_{[\alpha}B_{\beta]}^{ab}-\frac{1}{2}\delta_{\alpha}^{\mu}e_{a}^{\rho}e_{b}^{\sigma}\mathcal{D}_{[\rho}B_{\sigma]}^{ab}\right).\label{eq: EM_tot-simple}
\end{equation}

In this background, we again interested in the torsion-free condition
with the spin connection in Eq.(\ref{eq: spin_connection}). Therefore
the Friedmann equations in Eqs.(\ref{eq: Friedmann_00}), (\ref{eq: Friedmann_ii})
and (\ref{eq: Friedmann_acceleration}) reduce to the following forms,
\begin{equation}
\left(\frac{\dot{a}}{a}\right)^{2}=\frac{2\mu\dot{a}\psi}{a^{2}}+\Lambda,\label{eq: Friedmann_00_2}
\end{equation}

\begin{equation}
\frac{2\ddot{a}}{a}+\left(\frac{\dot{a}}{a}\right)^{2}=\frac{2\mu}{a^{2}}\left(\dot{a}\psi+a\dot{\psi}\right)+3\Lambda,\label{eq: Friedmann_ii_2}
\end{equation}

\begin{equation}
\frac{\ddot{a}}{a}=\frac{\mu\dot{\psi}}{a}+\Lambda,\label{eq: Friedmann_acceleration_2}
\end{equation}
and the energy density and the pressure expressions related to the
$B_{\mu}^{ab}\left(x\right)$ gauge field reduce to

\begin{eqnarray}
\rho_{B} & = & \frac{6\mu\dot{a}\psi}{\kappa a^{2}},\,\,\,\,\,\,\,P_{B}=-\frac{2\mu\dot{\psi}}{\kappa a^{2}}-\frac{1}{3}\rho_{B}.\label{eq: rho_P_B-m-1}
\end{eqnarray}
At this point, it should be noted that the second scalar field, denoted
as $\zeta\left(t\right)$, exhibits no influence on the Friedmann
equations under these particular circumstances. 

If we solve Eq.(\ref{eq: Friedmann_00_2}) with respect to $\psi\left(t\right)$,
we get
\begin{equation}
\psi=\frac{\dot{a}^{2}-\Lambda a^{2}}{2\mu\dot{a}},\label{eq: psi_3}
\end{equation}
then substituting this result to Eq.(\ref{eq: Friedmann_ii_2}) or
Eq.(\ref{eq: Friedmann_acceleration_2}), we find the following solutions
for the scale factor;
\begin{equation}
a=e^{\pm\sqrt{\Lambda}\left(t-C_{1}\right)},\,\,\,C_{1}t+C_{2},\label{eq: solution_a(t)}
\end{equation}
where $C_{1}$ and $C_{2}$ are the integration constants. For consistency,
one can use the solution of Eq.(\ref{eq: Friedmann_acceleration_2})
\begin{equation}
\psi=\int\frac{\ddot{a}-\Lambda a}{\mu}\text{d}t+C_{1},\label{eq: psi_4}
\end{equation}
and equate with Eq.(\ref{eq: psi_3}), this calculation leads to the
same solution as Eq.(\ref{eq: solution_a(t)}). The first solution
in Eq.(\ref{eq: solution_a(t)}) correspond to (anti) de Sitter universe
and this solution forces the function $\psi$ to be zero. On the other
hand, the second solution results in a non-accelerating universe due
to $\ddot{a}=0$. These findings suggest that the Maxwell gravity
formulation under consideration offers a limited range of solutions
and therefore lacks a diverse foundation for investigating the universe's
evolution.

\section{Conclusion\label{sec:Conclusion}}

In this paper, we analyzed the Maxwell gravity model which is based
on gauging the semi-simple extended Poincaré algebra \citep{soroka2012gauge}
in the context of cosmology. In analogy with the Einstein-Yang-Mills
theory, we described the Maxwell gauge field as a function of time
in terms of two time-dependent functions $\psi\left(t\right)$ and
$\zeta\left(t\right)$ in Eqs.(\ref{eq: B_1}) and (\ref{eq: B_2}).
By using this definition, we found the Friedmann equations involving
contributions of the Maxwell gauge field together with the cosmological
constant. Then we analyzed the solution of Friedman equations under
the exponential evolution of the scale factor and derived the explicit
expression of $\psi\left(t\right)$ and $\zeta\left(t\right)$. Thus
we show that this model provides a useful background to study one
of the most common cosmological models. In addition to this approach,
it is known that the power series evolution of the scale factor such
as $a\left(t\right)\approx t^{h}$ produces well-known cosmological
models such as the radiation dominated ($h=1/2$) or the matter dominated
($h=2/3$) universe. So one can study this model and derive the exact
expression of $\psi\left(t\right)$ and $\zeta\left(t\right)$. 

It is important to note that if one specifies $\psi\left(t\right)$
or $\zeta\left(t\right)$ at first instead of defining the scale factor,
this approach may lead to interesting cosmological models. On the
other hand, one can add a matter term to the action (\ref{eq: action})
instead of trying to describe the all matter-energy content of the
universe by the Maxwell curvature contributions to find more general
solutions. 

We also analyzed $\alpha=0$ condition which corresponds to the minimal
model of the Maxwell algebra given in \citep{soroka2005tensor,azcarraga2011generalized}
and we find that this model only allows the (anti) de Sitter universe
models and a non-accelerated universe model (see Eq.(\ref{eq: solution_a(t)})).
We also see that the function $\zeta\left(t\right)$ does not make
any contribution to the Friedmann equations for this condition. Despite
these limitations, we can easily say that by using different gravitational
actions (see \citep{azcarraga2011generalized}) one may derive different
cosmological models. 

Note that similar to the Maxwell algebra, the non-commutativity of
the momentum generators would lead to important experimental consequences
and a modification of the uncertainty principle (for the cosmological
effects, see \citep{Perivolaropoulos:2017Cosmological} and references
therein). If one establishes a connection between these studies and
the Maxwell algebra, this idea may produce interesting results or
gets some constraints or limits for the parameter $\lambda$.

Finally, we can say that the Maxwell symmetry offers a potential alternative
approach that may help the explanation of the evolution of our universe.
Furthermore, the additional gauge fields to the gravitation theory
presents a useful foundation to examine unresolved phenomena in General
Relativity, such as the inflationary universe (for more information
\citep{maleknejad2013gauge_report}) and dark energy models \citep{Zhao:2005TheStateEquation}.
Therefore, the model derived from this study will be explored elsewhere
in the context of these fields.

\section*{Conflicts of Interest }

The authors declare no conflicts of interest.
\begin{acknowledgments}
This study is supported by the Scientific and Technological Research
Council of Turkey (TUB\.{I}TAK) under grant number 2219.
\end{acknowledgments}

\bibliography{maxwell_cosmo}

\end{document}